\documentclass[aps,prl,twocolumn,superscriptaddress]{revtex4-1} 

\usepackage{graphicx}
\usepackage{dcolumn}
\usepackage{bm}

\begin{document}


\title{Quantum process tomography of two-qubit controlled-Z and controlled-NOT gates 
using superconducting phase qubits}


\author{T. Yamamoto}
\affiliation{Department of Physics, University of California, Santa Barbara, California 93106, USA}
\affiliation{Green Innovation Research Laboratories, NEC Corporation, Tsukuba, Ibaraki 305-8501, Japan}

\author{M. Neeley}
\affiliation{Department of Physics, University of California, Santa Barbara, California 93106, USA}

\author{E. Lucero}
\affiliation{Department of Physics, University of California, Santa Barbara, California 93106, USA}

\author{R. C. Bialczak}
\affiliation{Department of Physics, University of California, Santa Barbara, California 93106, USA}

\author{J. Kelly}
\affiliation{Department of Physics, University of California, Santa Barbara, California 93106, USA}

\author{M. Lenander}
\affiliation{Department of Physics, University of California, Santa Barbara, California 93106, USA}

\author{Matteo Mariantoni}
\affiliation{Department of Physics, University of California, Santa Barbara, California 93106, USA}

\author{A. D. O'Connell}
\affiliation{Department of Physics, University of California, Santa Barbara, California 93106, USA}

\author{D. Sank}
\affiliation{Department of Physics, University of California, Santa Barbara, California 93106, USA}

\author{H. Wang}
\affiliation{Department of Physics, University of California, Santa Barbara, California 93106, USA}

\author{M. Weides}
\affiliation{Department of Physics, University of California, Santa Barbara, California 93106, USA}

\author{J. Wenner}
\affiliation{Department of Physics, University of California, Santa Barbara, California 93106, USA}

\author{Y. Yin}
\affiliation{Department of Physics, University of California, Santa Barbara, California 93106, USA}

\author{A. N. Cleland}
\email[]{anc@physics.ucsb.edu}
\affiliation{Department of Physics, University of California, Santa Barbara, California 93106, USA}

\author{John M. Martinis}
\email[]{martinis@physics.ucsb.edu}
\affiliation{Department of Physics, University of California, Santa Barbara, California 93106, USA}


\date{\today}

\begin{abstract}
We experimentally demonstrate quantum process tomography of controlled-Z and 
controlled-NOT gates using capacitively-coupled superconducting phase qubits. 
These gates are realized by using the $|2\rangle$ state of the phase qubit. 
We obtain a process fidelity of 0.70 for the controlled-phase and 0.56 for the controlled-NOT gate, 
with the loss of fidelity mostly due to single-qubit decoherence. 
The controlled-Z gate is also used to demonstrate a two-qubit Deutsch-Jozsa algorithm with a single function query.
\end{abstract}

\pacs{}

\maketitle

Quantum computation and quantum communication rely on excellent control of the underlying quantum system~\cite{Ladd10}. Reasonable control has been achieved with a variety of quantum systems, with superconducting qubits emerging as one of the most promising candidates~\cite{Clarke08}.  Recent experiments using superconducting architectures include demonstrations of quantum algorithms using two qubits~\cite{DiCarlo09} and the entanglement of three qubits~\cite{Neeley10,DiCarlo10}.  A key element in these experiments is a two-qubit entangling gate, such as the $\sqrt{i\rm SWAP}$ \cite{Neeley10} and the controlled-Z (CZ) gates~\cite{DiCarlo09,DiCarlo10}.  Because the CZ gate is simple to implement, has high fidelity, and can readily generate controlled-NOT (CNOT) logic~\cite{NielsenBook}, it likely will be an important component in more complex algorithms such as quantum error correction.  At present, however, the CZ gate functionality has only been directly tested for a subset of the possible input states.

In this Letter, we demonstrate the operation of a CZ gate in a superconducting phase qubit, and fully characterize this gate as well as a CNOT gate 
using quantum process tomography (QPT). We additionally use the CZ gate to perform the Deutsch-Jozsa algorithm~\cite{DiCarlo09}, here with a single-shot evaluation of the function. The use of QPT provides a more complete gate evaluation than, for example, measuring the truth table for the corresponding CNOT gate~\cite{Yamamoto03,Plantenberg07}, as it verifies that the gate will properly transform any possible input state.  QPT for two- or three-qubit gates has been reported in NMR~\cite{Childs01}, optics~\cite{Obrien04,Langford05,Kiesel05}, and in ion traps~\cite{Riebe06,Monz09}.  In solid state systems, QPT has been implemented for the $\sqrt{i\rm SWAP}$ gate with the phase qubit~\cite{Bialczak10}.

\begin{figure}[b]
\includegraphics[width=0.9\columnwidth,clip]{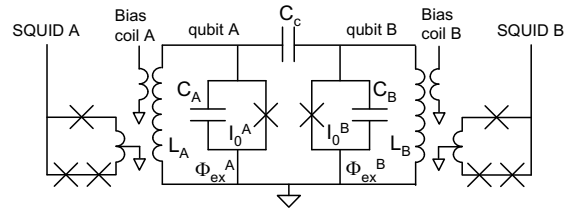}
\caption{~\label{Fig1}
Circuit diagram for the experimental device, showing two flux-biased phase qubits coupled by a fixed capacitance $C_c$.  A bias coil and readout SQUID are coupled to each qubit.  The design parameters of the circuit are $I_0^{\rm A}=I_0^{\rm B}=2~\mu$A, $C_{\rm A}=C_{\rm B}=1$~pF,
$L_{\rm A}=L_{\rm B}=720$~pH, and $C_{\rm c}=2$~fF.
}
\end{figure}

The electrical circuit for the device is shown in Fig.~\ref{Fig1}, comprising two superconducting phase qubits A and B, coupled by a fixed capacitance $C_{\rm c}$.  Each qubit is a superconducting loop interrupted by a capacitively-shunted Josephson junction.
When biased close to the critical current, the junction and its parallel loop inductance produce a non-linear potential as a function of the phase difference across the junction. Combined with the kinetic energy originating from the shunting capacitance, unequally spaced quantized energy levels appear in the cubic potential. The two lowest levels are used for the qubit states $|0\rangle$ and $|1\rangle$, with a transition frequency $f_{\rm 10}^{\rm A}$ ($f_{\rm 10}^{\rm B}$) that can be controlled by an external magnetic flux $\Phi_{\rm ex}^{\rm A}$ ($\Phi_{\rm ex}^{\rm B}$) applied to the loop.  The third energy level $|2\rangle$ is used as an auxiliary state to realize the CZ gate, as discussed below.

The operation of a similar device has been reported previously \cite{Steffen06b,Bialczak10}. The state of each qubit is controlled by applying a rectangular-shaped current pulse (Z-pulse) or a Gaussian-shaped microwave pulse (X-, Y-pulse) to its bias coil.  For an X- or Y-pulse, we simultaneously apply the derivative of the pulse to the quadrature (90${}^\circ$ phase shifted) drive to reduce both unwanted excitation of the $|2\rangle$ state and phase error due to AC Stark effect~\cite{Motzoi09}; the derivative scaling factor is determined from the nonlinearity of each qubit~\cite{Lucero10}.  This procedure enables us to use a Gaussian pulse with a full-width at half maximum (FWHM) of 10~ns, while maintaining accurate qubit control~\cite{Lucero08} in spite of a  rather weak qubit nonlinearity ($\sim 100$~MHz).  Each qubit state is read out individually in a single-shot manner by injecting a large magnitude Z-pulse and then measuring the qubit flux with a superconducting quantum interference device (SQUID).

The device was fabricated using a photolithographic process with Al films, AlO$_{x}$ tunnel junctions, and a-Si:H dielectric for the shunt capacitors and wiring crossovers, all on a sapphire substrate.  The device was mounted in a superconducting aluminum sample holder and cooled in a dilution refrigerator to $\sim25$~mK.

In the present experiment, the two qubits were biased so that
$f_{\rm 10}^{\rm A}=7.16$~GHz and $f_{\rm 10}^{\rm B}=7.36$~GHz when no Z-pulse was applied.  The relaxation times ($T_1$) were measured to be 510 ns and 500 ns for qubit A and B, respectively.  The dephasing times determined from a Ramsey interference experiment ($T_2^{\rm Ramsey}$), which showed Gaussian decay proportional to $\exp{[-(t/T_2^{\rm Ramsey})^2]}$ due to $1/f$ flux noise ~\cite{Bialczak07}, were 200 and 230 ns, respectively.

\begin{figure}[t]
\includegraphics[width=0.9\columnwidth,clip]{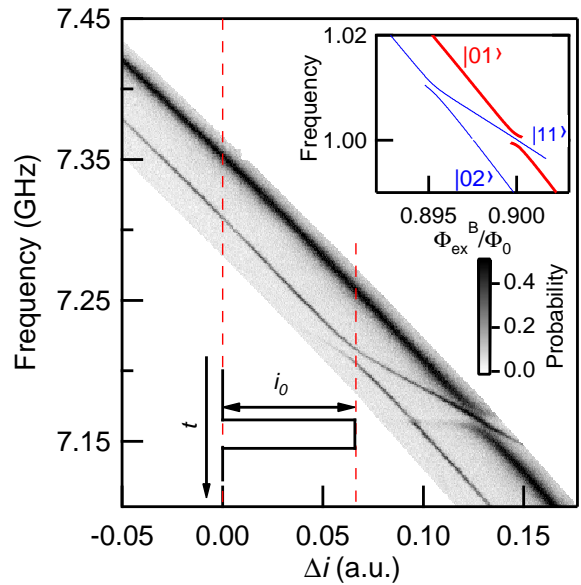}
\caption{~\label{Fig2}~(Color online)
High-power spectroscopy for qubit B. The escape probability (gray scale) is plotted versus microwave frequency and the Z-pulse amplitude $\Delta i$.  The single photon $|0\rangle \rightarrow |1\rangle$ and two-photon  $|0\rangle \rightarrow |2\rangle$ transitions are visible, along with two avoided-level crossings.  The upper inset shows the calculated states and eigenenergies, with the thick (thin) lines representing single (two) photon excitations. The lower inset illustrates the Z-pulse amplitude $i_0$ for the $\textrm{CZ} = -(\textrm{SWAP})^2$ operation.
}
\end{figure}

Figure~\ref{Fig2} shows the high-power spectroscopy for qubit B, which is used to guide formation of the CZ gate.  We plot the escape probability of qubit B in gray scale as a function of the amplitude $\Delta i$ of a 2~$\mu$s long Z-pulse (horizontal axis) and the frequency of a microwave X-pulse (vertical axis) of the same length.  Both pulses were applied simultaneously to qubit B, followed by the Z-pulse for the readout. 
In this way, we can probe the change in the resonance frequency $f_{\rm 10}^{\rm B}$ as a function of detuning $\Delta i$.  In addition to the main resonance line corresponding to $f_{\rm 10}^{\rm B}$, somewhat broadened because of the large amplitude of the microwave pulse ($\Delta\Phi_{\rm ex}^{\rm B} \sim 10~\mu\Phi_0$), a sharper line is observed on the low-frequency side of the main resonance; this corresponds to the two-photon excitation from the $|0\rangle$ to the $|2\rangle$ state~\cite{Bushev10}.  The vertical distance between the main and two-photon lines is 1/2 the qubit nonlinearity  $\Delta f = f_{\rm 10} - f_{\rm 21}$, yielding $\Delta f=$  114 MHz for qubit A (data not shown) and 87 MHz for qubit B.

We observe an avoided level crossing in the main resonance at $\Delta i \simeq 0.13$ when the two qubit frequencies overlap $f_{\rm 10}^{\rm B} = f_{\rm 10}^{\rm A}$.  Here, the degeneracy of the $|AB\rangle = |10\rangle$ and $|01\rangle$ states produces a splitting with size 14.2$\pm$0.2~MHz, determined from a fit to the data, consistent with the designed capacitance $C_{\rm c}$.  The avoided crossing for the two-photon line at $\Delta i \simeq 0.066$ gives a splitting of 9.7$\pm$0.2~MHz, about $\sqrt{2}/2$ times as large as the main resonance, as expected from a $|11\rangle$ and $|02\rangle$ interaction.  The slope of the resonance between the two crossings is 1/2 that of $f_{10}^B$, as expected for the $|11\rangle$ state.

Our interpretation of the spectroscopy is validated by a numerical calculation.  Using three states for each qubit and the qubit design parameters, we calculate from the resulting $9 \times 9$ Hamiltonian~\cite{Steffen03,Kofman07} the energies for the coupled eigenstates.
The energy bands, normalized to $f_{\rm 01}^{\rm A}$, are plotted versus the flux bias for qubit B in the upper inset of Fig.~\ref{Fig2}.  Here, a band is plotted only when its transition matrix element from the ground state is above a threshold, to simulate the appearance of the transition in the spectroscopic measurement~\cite{suppl}.  The (red) thick lines correspond to the $|01\rangle$ and $|10\rangle$ states, whereas the (blue) thin lines represent half of the excitation energy of the $|11\rangle$ and $|02\rangle$ states.  The overall structure agrees well with the experimental data.

As proposed theoretically by Strauch {\it et al.}~\cite{Strauch03},
the avoided crossing due to the degeneracy of the $|11\rangle$ and
$|02\rangle$ states can be used to construct a CZ gate, whose action produces no change in state except for $|11\rangle \rightarrow -|11\rangle$.
By applying a non-adiabatic Z-pulse,  the $|11\rangle$ and $|02\rangle$ states become degenerate (see lower inset of Fig.~\ref{Fig2}).
Initially in the $|11\rangle$ state, the system evolves as an  $i\textrm{SWAP}$ interaction, giving
$|\Psi(t)\rangle = \cos(\gamma \Delta t/\hbar)|11\rangle + i\sin(\gamma \Delta t/\hbar)|02\rangle$, where $2\gamma$ is the splitting energy of the avoided crossing and $\Delta t$ the duration of the Z-pulse.  After twice the $i\textrm{SWAP}$ time $\Delta t$ = $h/2\gamma$, the system returns to the initial state $|11\rangle$, but with a minus sign.  If the system starts in $|00\rangle$, $|01\rangle$, or $|10\rangle$, the state does not change since it is off-resonance with both avoided level crossings.  
A similar scheme using an adiabatic Z pulse has been used to successfully demonstrate a quantum algorithm~\cite{DiCarlo09}, 
and the same (non-adiabatic) scheme has recently been used to create a three qubit entangled state in transmon qubits~\cite{DiCarlo10}.

\begin{figure}[t]
\includegraphics[width=1.0\columnwidth,clip]{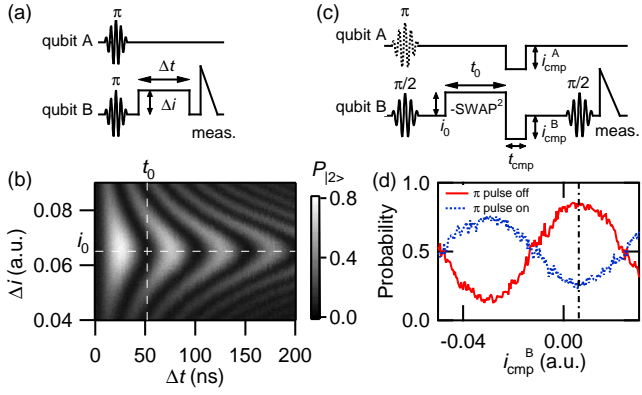}
\caption{~\label{Fig3}~(Color online)
(a) Operation sequence for (b), the state evolution between $|11\rangle$ and $|02\rangle$ states.
(b) Plot of $|2\rangle$ state probability of qubit B versus Z-pulse time $\Delta t$ and Z-pulse amplitude $\Delta i$.  The dashed lines correspond to the optimal setting for the CZ gate.
(c) Operation sequence for (d), demonstration of the CZ gate.
(d) Plot of $|1\rangle$ state probability of qubit B as a function of $i_{\rm cmp}^B$ ($i_{\rm cmp}^A$ is fixed as 3 $\times 10^{-4}$.). The (red) solid and (blue) dashed curves are for qubit A initialized to the $|0\rangle$ and $|1\rangle$ states, respectively.  The vertical dot-dashed line indicates the value of $i_{\rm cmp}^B$ for the CZ gate.
}
\end{figure}

To experimentally determine the amplitude and length of the required non-adiabatic Z-pulse, we directly measured the coherent oscillation between the $|11\rangle$ and $|02\rangle$ states.  This $(i\textrm{SWAP})^2$ operation sequence is shown in Fig.~\ref{Fig3}(a): We first prepare the $|11\rangle$ state with a $\pi$ pulse to both qubits, and then apply a Z-pulse with amplitude  $\Delta i$ and length $\Delta t$ to qubit B.  
Here only qubit B is probed and we adjust the measurement pulse amplitude so that the qubit is detected only when in the $|2\rangle$ (or higher) state~\cite{Neeley09}.  In Fig.~\ref{Fig3}(b), we plot the tunneling probability $P_{|2\rangle}$  as a function of $\Delta t$ and $\Delta i$, which shows the expected chevron pattern.  The minimum oscillation frequency occurs at a value of $\Delta i$ that agrees with $i_0$ determined in Fig.~\ref{Fig2}(a).  The oscillation period $t_0 = 51.8\,\textrm{ns}$ is also consistent with the splitting size of the avoided crossing.  At the intersection of these two dashed lines, the time evolution of the state produces a minus sign, as required for the CZ gate. We stress that no discernable increase in $P_{|2\rangle}$ is observed ($<1\%$) at this operation point $(\Delta t, \Delta i)= (t_0,i_0)$, confirming that we return to the $|11\rangle$ state after the CZ operation.


Because the qubits themselves also accumulate phase $\phi$ during the CZ pulse, the general unitary evolution from the gate is given by
\begin{equation}~\label{cphase}
U = \left(
\begin{array}{cccc}
1 & 0 & 0 & 0 \\
0 & e^{i\phi_A} & 0 & 0 \\
0 & 0 & e^{i\phi_B} & 0 \\
0 & 0 & 0 & -e^{i(\phi_A+\phi_B)} \\
\end{array}
\right).
\end{equation}
By adding additional Z-pulses to both qubits, we can compensate these phases and even place the minus sign at any diagonal position in the matrix~\cite{DiCarlo09}.  The compensation pulses are shown in Fig.~\ref{Fig3}(c), which consist of a fixed 10~ns pulse of variable amplitude $i_{\rm cmp}$ after the CZ pulse.
In Fig.~\ref{Fig3}(d), we plot the tunneling probability of qubit B
as a function of $i_{\rm cmp}^B$ for fixed $i_{\rm cmp}^A$.  The phase of qubit B is measured through a Ramsey fringe experiment. The (red) solid and (blue) dashed curves correspond to qubit A being in the $|0\rangle$ or $|1\rangle$ state.  They both show a sinusoidal dependence on $i_{\rm cmp}^B$, but are shifted by $\pi$ from each other, confirming the correct operation of the CZ gate.  A similar experiment was done for qubit A (data not shown).

The phases for the CZ gate are set by taking the values of $i_{\rm cmp}$ that give maximum probability when the control qubit is in the $|0\rangle$ state, as indicated by the vertical dash-dotted line in Fig.~\ref{Fig3}(d).  Controlled-NOT (CNOT) gates are constructed
by combining the CZ gate with single qubit rotations
$U_{\rm CNOT} = (I \otimes R^{\pi/2}_y)~CZ~(I \otimes R^{-\pi/2}_y)$, where $R^{\theta}_y$ represents the rotation of a single qubit state by an angle $\theta$ about the $y$ axis, and $I$ is the identity operator.


\begin{figure}[ttt]
\includegraphics[width=1.0\columnwidth,clip]{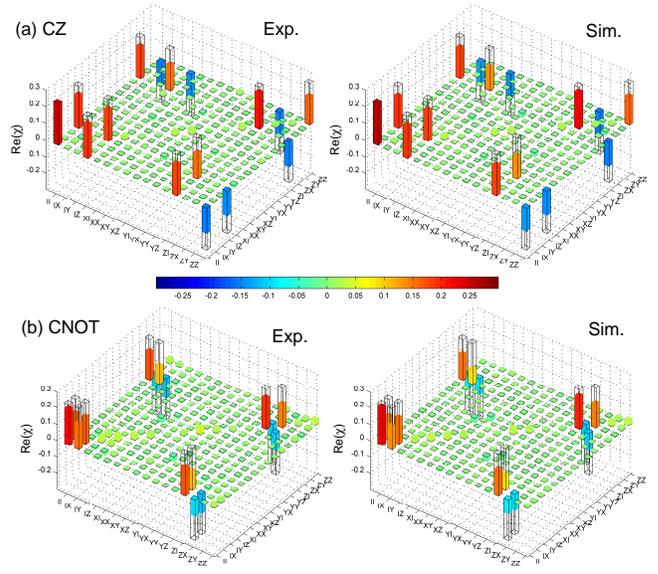}
\caption{~\label{Fig4}~(Color online)
$\chi$ matrices of CZ and CNOT gates. 
(a) Left panel: The real part of the experimentally obtained $\chi$ matrix ($\chi_{\rm p}$) for the CZ gate, with $F_{\rm p} = 0.70$.
Right panel: The real part of the simulated $\chi$ matrix for the CZ gate, with $F_{\rm p} = 0.67$.
(b) Left panel: The real part of $\chi_{\rm p}$ for the CNOT gate, with $F_{\rm p} = 0.56$. 
Right panel: The real part of the simulated $\chi$ matrix for the CNOT gate, with $F_{\rm p} = 0.52$.
The open boxes in the figure represent the ideal $\chi$ matrix.
}
\end{figure}

We evaluate the performance of these gates with QPT. 
For QPT, we prepare 16 input states in total, chosen from the set
$\{|0\rangle, |1\rangle, |0\rangle + |1\rangle, |0\rangle + i|1\rangle \}$ for each qubit.  After preparing these input states, we determine the density matrix of the output state with quantum state  tomography~\cite{Steffen06b}, in which we measure each qubit along the six directions $\pm x$, $\pm y$ and $\pm z$ of the Bloch sphere~\cite{Neeley08}.  For each combination of QPT and QST pulses, we repeat the sequence 1800 times to obtain the joint qubit probabilities $P_{AB}=P_{00}$,$P_{10}$,$P_{01}$ and $P_{11}$. After correcting for small measurement errors~\cite{Bialczak10}, we reconstruct the $16 \times 16$ experimental $\chi_{\rm e}$ matrix from the resulting 16 density matrices~\cite{comment1}.  
With experimental noise, the $\chi_{\rm e}$ matrix found in this way is not necessarily physical, {\it i.e.} completely positive and trace-preserving.
We thus use convex optimization to obtain the physical matrix $\chi_{\rm p}$ 
that best approximates $\chi_{\rm e}$, as used in Ref.~\cite{Bialczak10}.

We plot the real part of $\chi_{\rm p}$ for the CZ and CNOT gates
in the left panel of Figs.~\ref{Fig4}(a) and (b). The results are displayed in the basis formed by the Kronecker product of Pauli operators $\{I, \sigma_x, -i\sigma_y, \sigma_z \}$ for each qubit~\cite{NielsenBook}. The open boxes represent the ideal $\chi$ matrix.  The imaginary parts of $\chi_{\rm p}$ have very small magnitude ($<0.04$ for CZ and $<0.03$ for CNOT), and are shown in Ref.~\cite{suppl}.
For both gates, we observe elements with large amplitudes at the proper positions.
More quantitative evaluation is obtained by calculating the process fidelity $F_{\rm p}$, defined by $F_{\rm p}={\rm Tr}(\chi_{\rm i}\chi_{\rm p})$, where $\chi_{\rm i}$ represents an ideal $\chi$ matrix. We obtain $F_{\rm p}=0.70$ for the experimentally measured CZ gate, and 0.56 for the CNOT gate.
For CZ gates with a minus sign at other positions on the diagonal, the measured $F_{\rm p}$'s are 0.68, 0.69, and 0.70 for CZ$_{00}$, CZ$_{01}$, and CZ$_{10}$, respectively~\cite{suppl}.

To understand the loss of process fidelity, we performed numerical simulations by solving the master equation using the experimental parameters~\cite{suppl}.
The entire sequence, including the QPT and QST pulses, was simulated to construct $\chi_{\rm sim}$.  The real part of $\chi_{\rm sim}$ is shown in Fig.~\ref{Fig4}, and reproduces reasonably well the reduction of the expected elements and the appearance of small unwanted elements.
These imperfections are removed as we increase the single-qubit coherence time in the simulation, which suggests that loss of $F_{\rm p}$ in our system is mostly dominated by single qubit decoherence.
We note that it is possible to obtain more information on the decoherence mechanisms
by analyzing the magnitude of particular elements in the $\chi$ matrix~\cite{Kofman09}.

By using these conditional gates, we can perform the Deutsch-Jozsa algorithm~\cite{NielsenBook} using the pulse sequence described in Ref.~\cite{DiCarlo09}.  The experimental probability to obtain the correct answer is summarized in Table~\ref{table1}.  Because our phase qubit has single-shot readout, we can obtain the correct answer to a single function query more than 70$\%$ of the time, greater than the 50$\%$ probability for a classical query and guess.
We stress that no calibration for the measurement error is applied here.
The full density matrix of the final state is given in Ref.~\cite{suppl}.

\begin{table}[t]
\caption{\label{table1}%
Summary of performance for Deutsch-Jozsa algorithm.
Deutsch-Jozsa functions are defined as
$f_0(x)=0$, $f_1(x)=1$, $f_2(x)=x$, and $f_3(x)=1-x$.}
\begin{ruledtabular}
\begin{tabular}{cccccc}
\multicolumn{2}{c}{\textrm{Element}} &
\multicolumn{4}{c}{\textrm{Deutsch-Jozsa function}} \\
 & & \multicolumn{2}{c}{\textrm{Constant}} & \multicolumn{2}{c}{\textrm{Balanced}} \\
 & & $f_0$ & $f_1$ & $f_2$ & $f_3$ \\
\colrule
$\langle 00| \rho |00 \rangle + \langle 01| \rho |01 \rangle$ & ideal & 0 & 0 & 1 & 1 \\
 & measured & 0.29 & 0.28 & 0.76 & 0.74 \\
$\langle 10| \rho |10 \rangle + \langle 11| \rho |11 \rangle$ & ideal & 1 & 1 & 0 & 0 \\
 & measured & 0.71 & 0.72 & 0.24 & 0.26
\end{tabular}
\end{ruledtabular}
\end{table}

In conclusion, we have demonstrated CZ and CNOT gates in capacitively-coupled phase qubits using the higher-energy $|2\rangle$ state.  Quantum process tomography measures a $\chi$ matrix that is in good accord with predictions, which is a definitive test of proper gate operation for any input state.

\begin{acknowledgments}
The authors would like to thank F. K. Wilhelm and A. N. Korotkov for valuable discussion. 
They would also like to thank Y. Nakamura for useful comments on the manuscript. 
M. M. acknowledges support from an Elings fellowship. 
Semidefinite programming convex optimization was carried out using
the open-source MATLAB packages YALMIP and SeDuMi.  This work was supported by
IARPA under ARO award W911NF-04-1-0204.
\end{acknowledgments}


\begin{thebibliography}{29}
\bibitem [{com()}]{comment1}%
  \BibitemOpen
  \href@noop {} {}\bibinfo {note} {We assumed that input states are ideally
  prepared.}\BibitemShut {Stop}%
\end{thebibliography}

%

\end{document}



\title{Quantum process tomography of two-qubit controlled-Z and controlled-NOT gates 
using superconducting phase qubits: Supplementary information}


\author{T. Yamamoto}
\affiliation{Department of Physics, University of California, Santa Barbara, California 93106, USA}
\affiliation{Green Innovation Research Laboratories, NEC Corporation, Tsukuba, Ibaraki 305-8501, Japan}

\author{M. Neeley}
\affiliation{Department of Physics, University of California, Santa Barbara, California 93106, USA}

\author{E. Lucero}
\affiliation{Department of Physics, University of California, Santa Barbara, California 93106, USA}

\author{R. C. Bialczak}
\affiliation{Department of Physics, University of California, Santa Barbara, California 93106, USA}

\author{J. Kelly}
\affiliation{Department of Physics, University of California, Santa Barbara, California 93106, USA}

\author{M. Lenander}
\affiliation{Department of Physics, University of California, Santa Barbara, California 93106, USA}

\author{Matteo Mariantoni}
\affiliation{Department of Physics, University of California, Santa Barbara, California 93106, USA}

\author{A. D. O'Connell}
\affiliation{Department of Physics, University of California, Santa Barbara, California 93106, USA}

\author{D. Sank}
\affiliation{Department of Physics, University of California, Santa Barbara, California 93106, USA}

\author{H. Wang}
\affiliation{Department of Physics, University of California, Santa Barbara, California 93106, USA}

\author{M. Weides}
\affiliation{Department of Physics, University of California, Santa Barbara, California 93106, USA}

\author{J. Wenner}
\affiliation{Department of Physics, University of California, Santa Barbara, California 93106, USA}

\author{Y. Yin}
\affiliation{Department of Physics, University of California, Santa Barbara, California 93106, USA}

\author{A. N. Cleland}
\affiliation{Department of Physics, University of California, Santa Barbara, California 93106, USA}

\author{John M. Martinis}
\affiliation{Department of Physics, University of California, Santa Barbara, California 93106, USA}


\date{\today}

\pacs{}

\maketitle

\section{Calculation of the energy bands and transition matrix elements}
We calculated the energy band of the capacitively-coupled flux-biased phase qubits
by diagonalizing the following $9 \times 9$ Hamiltonian~\cite{Steffen03,Kofman07}, 
\begin{widetext}
\begin{equation}~\label{3level_hami}
\mathcal{H} \simeq
\left(
\begin{array}{ccc}
0 & &  \\
 & hf_{10}^{(A)} &  \\
 & & hf_{10}^{(A)} + hf_{21}^{(A)}
\end{array}
\right) \otimes I_2 + I_1 \otimes
\left(
\begin{array}{ccc}
0 & &  \\
 & hf_{10}^{(B)} &  \\
 & & hf_{10}^{(B)} + hf_{21}^{(B)}
\end{array}
\right) -g
\left(
\begin{array}{ccc}
0 & -1 & 0 \\
1 & 0 & -\sqrt{2} \\
0 & \sqrt{2} & 0
\end{array}
\right)
\otimes
\left(
\begin{array}{ccc}
0 & -1 & 0 \\
1 & 0 & -\sqrt{2} \\
0 & \sqrt{2} & 0
\end{array}
\right),
\end{equation}
\end{widetext}
where $f_{i,j}^{(A)}$ ($f_{i,j}^{(B)}$) is the flux-dependent transition frequency between $i$th and $j$th state of the qubit A (B),
and $g$ is the coupling energy between the qubits. 
The last term in the Hamiltonian is based on $\sigma_y \sigma_y$-type coupling of the two qubits. 
To calculate the transition matrix from the ground state $|g \rangle$ to
the excited state $|e \rangle$ in the spectroscopy experiment,
we calculated the transition matrix element of
$|\langle e|a^{\dagger} + a|g\rangle|^2$ for one photon excitation and
$\displaystyle \Bigl|\sum_i\frac{\langle e|a^{\dagger} + a|i\rangle \langle i|a^{\dagger} + a|g\rangle}{E_{e}-E_i-h f_{\rm d}}\Bigr|^2$
for two-photon excitation~\cite{CohenBook}, where $a$ ($a^{\dagger}$) is an annihilation (creation) operator for the harmonic oscillator,
$E_{i}$ is the energy gap of the state $|i\rangle$ from the ground state, and
$f_{\rm d}$ is the frequency of the $\mu$-wave drive,
which is set to be $E_{e}/2h$ in the calculation.

\section{$\chi$ matrix for all gates}
\begin{figure} 
\includegraphics[width=0.6\columnwidth,clip]{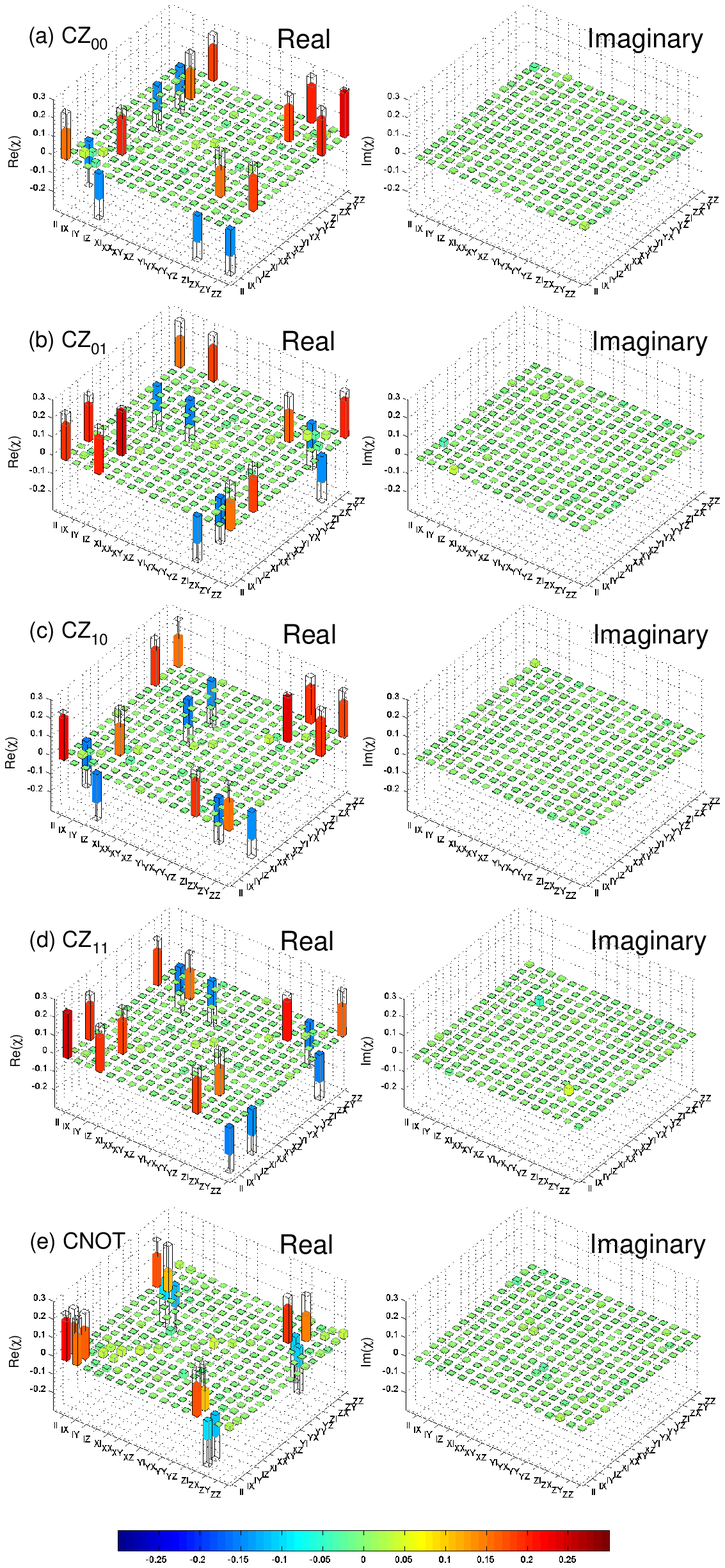}
\caption{~\label{FigS1}
$\chi_{\rm p}$ of (a) CZ$_{00}$ (b) CZ$_{01}$ (c) CZ$_{10}$ (d) $\textrm{CZ}=\textrm{CZ}_{11}$ and (e) CNOT.
Process fidelity $F_{\rm p}$ are 0.68, 0.69, 0.70, 0.70 and 0.56, respectively.
}
\end{figure}

In Fig.~\ref{FigS1}, the physical $\chi$ matrix $\chi_{\rm p}$ is plotted for all the CZ and CNOT gates.

\section{The difference between $\chi_{\rm p}$ and $\chi_{\rm e}$}
\begin{figure} [ttt]
\includegraphics[width=0.5\columnwidth,clip]{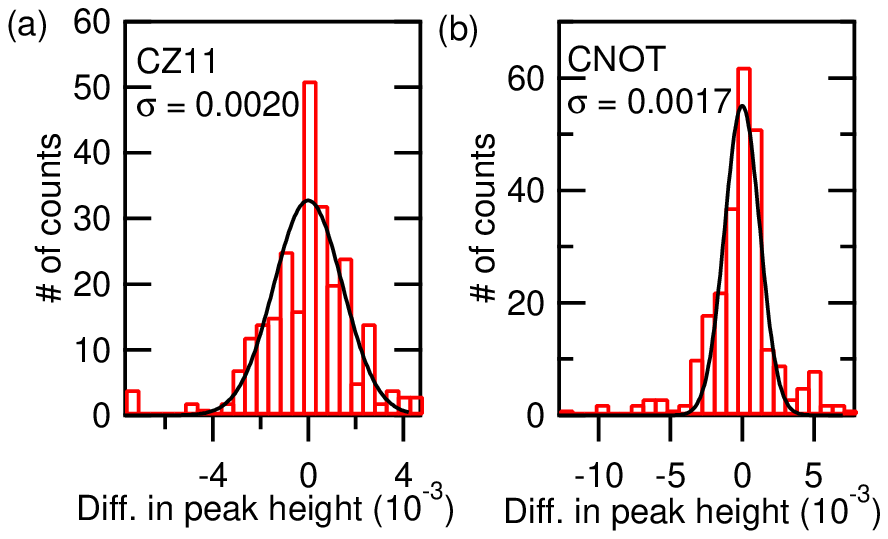}
\caption{~\label{FigS2}
Histogram of the differences in the peak height
of each of the 256 matrix elements in the real part of $\chi$ matrix.  
Data is for (a) CZ and (b) CNOT. The solid curves are a Gaussian fit to
the data.
}
\end{figure}

We checked the difference between $\chi_{\rm e}$ (the experimental $\chi$ matirix) and 
$\chi_{\rm p}$ (the physical $\chi$ matirix) by histogramming the differences in the peak height
$\Delta = \chi_{\rm e} - \chi_{\rm p}$
of each of the 256 matrix elements in the real part~\cite{Obrien04}.
We fit it by Gaussian $a\exp(-\Delta^2/\sigma^2)$ as shown in Fig.~\ref{FigS2}. The obtained $\sigma$ are 0.0020 for CP$_{11}$ and 0.0017 for CNOT gate,
which implies $\chi_{\rm e}$ and $\chi_{\rm p}$ are close.

\section{Simulation of QPT}
\begin{figure}
\includegraphics[width=0.6\columnwidth,clip]{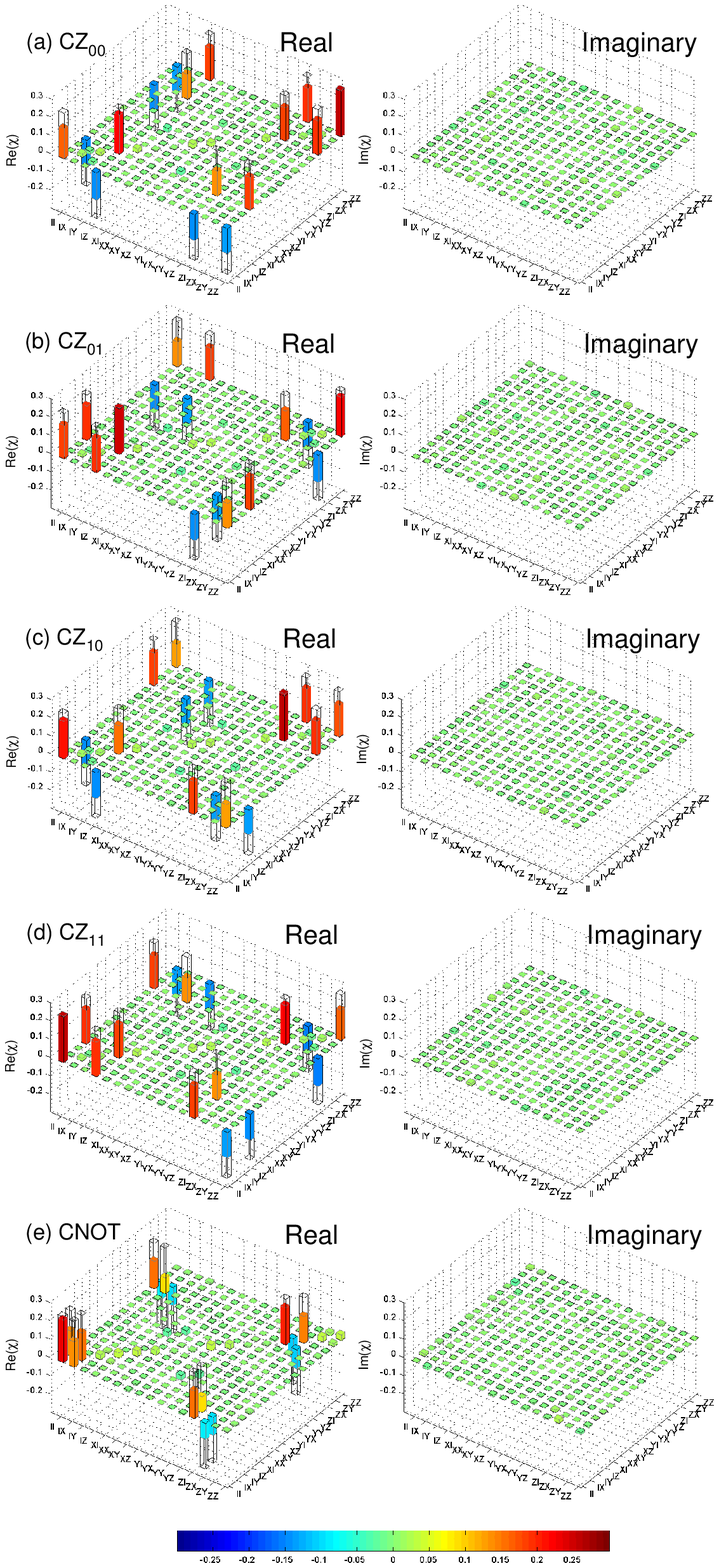}
\caption{~\label{FigS3}
Simulated $\chi$ matrix of (a) CZ$_{00}$ (b) CZ$_{01}$ (c) CZ$_{10}$ (d) CZ and (e) CNOT.
Process fidelity $F_{\rm p}$ are 0.67, 0.67, 0.66, 0.67 and 0.52, respectively.
}
\end{figure}

To simulate QPT, we solved the standard master equation,
$\dot{\rho}=-(i/\hbar)[\mathcal{H},\rho] + \mathcal{L}[\rho]$,
where $\mathcal{H}$ is a 9 by 9 Hamiltonian for capacitively coupled phase qubits under rotating wave approximation, and
$\displaystyle \mathcal{L}[\rho] = \sum_{i={\rm A,B}}\sum_{j=1,2} \mathcal{L}_j^i \rho \mathcal{L}_j^{i~\dagger} -
\frac{1}{2}\mathcal{L}_j^{i~\dagger} \mathcal{L}_j^i \rho - \frac{1}{2}\rho \mathcal{L}_j^{i~\dagger} \mathcal{L}_j^i$.
Here, for example, $\mathcal{L}_1^{\rm A} = a_{\rm A}/\sqrt{T_1^{\rm A}}$ and $\mathcal{L}_2^{\rm A} =
a_{\rm A}^\dagger a_{\rm A}\sqrt{2/T_2^{\rm A}}$ describe the relaxation and dephasing
for qubit A, respectively~\cite{Walls85}.
Experimental Ramsey interference shows Gaussian decay,
which is not reproduced by the above master equation.
Thus, in order to approximate this situation, we used an effective $T_2$ that
depends on the length of the control sequence for a particular experiment $t_{\rm seq}$.  In particular, we used $T_2 = {T_2^{\rm Ramsey}}^2/t_{\rm seq}$ in the simulation in order for both the Gaussian decay and exponential decay to give the same decay factor at $t_{\rm seq}$.
The actual $t_{\rm seq}$ is 101.8~ns in QPT for CZ gates and 141.8~ns for CNOT gate.
In Fig.~\ref{FigS3}, we show the simulated $\chi$ matrix of all CZ and CNOT gates.

\section{Experiment on Deutsch-Jozsa algorithm}
\begin{figure}
\includegraphics[width=0.5\columnwidth,clip]{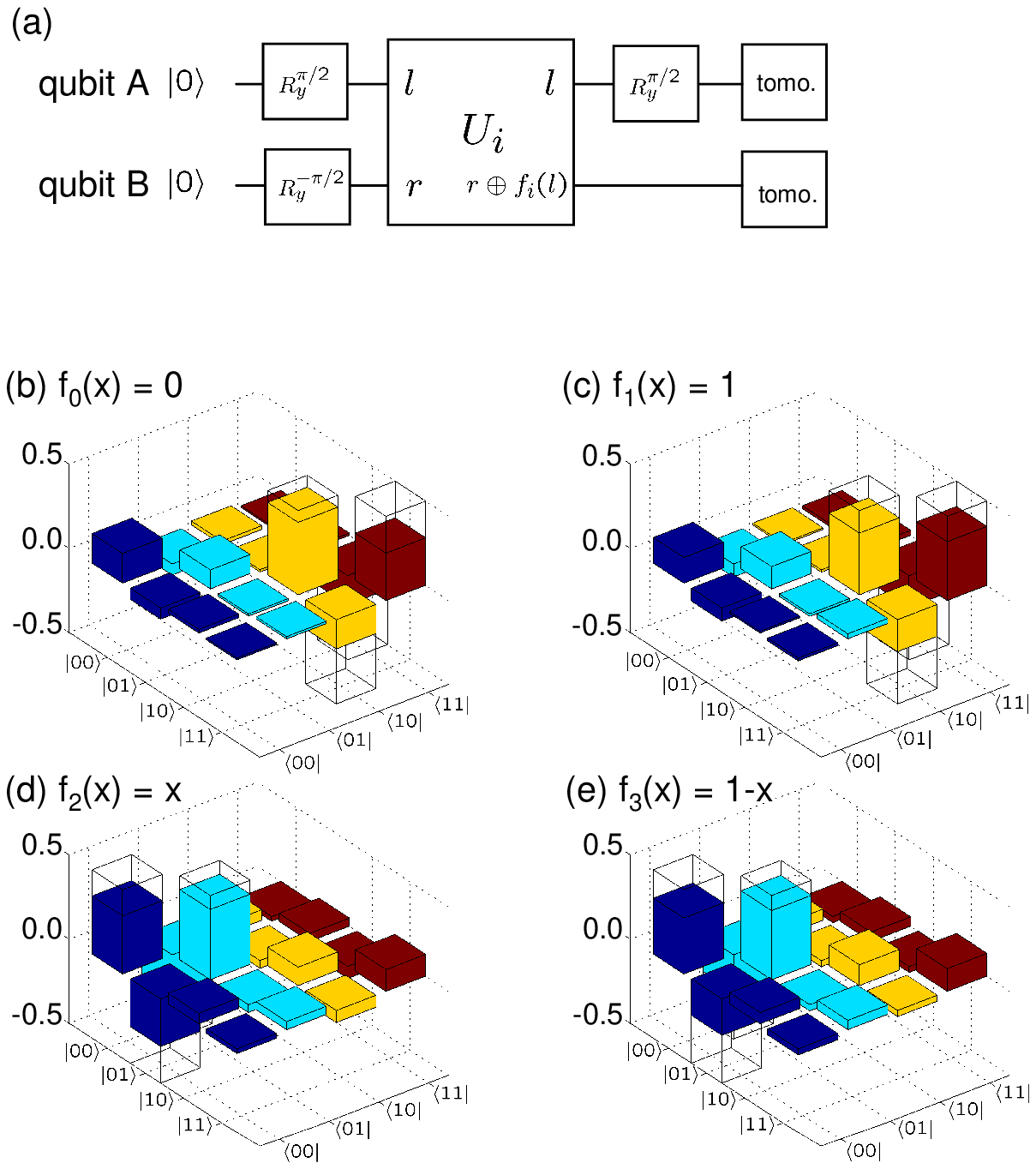}
\caption{~\label{FigS4}
(a) Pulse sequence for the Deutsch-Jozsa algorithm.
(b)-(d) Real part of the density matrix of the final state for four Deutsch-Jozsa functions. 
Open boxes represent the ideal density matrix. }
\end{figure}

Figure~\ref{FigS4}(a) shows the pulse sequence for the Deutsch-Jozsa algorithm. The four different two-qubit gates $U_i$ correspond to the four Deutsch-Jozsa
functions, which we want to determine by a single quantum evaluation of the function.
They are given by
\begin{equation}
  \begin{array}{l}
    U_0 = I \otimes I \ ,\\
    U_1 = I \otimes R_x^\pi \ ,\\
    U_2 = (I \otimes R_y^{\pi/2} R_x^\pi)\,CZ_{00}\,(I \otimes R_y^{\pi/2}) \ ,\\
    U_3 = (I \otimes R_y^{-\pi/2} R_x^\pi)\,CZ_{11}\,(I \otimes R_y^{-\pi/2}) \ .\\
  \end{array}
\end{equation}
The sequence is same as that used in Ref.~\cite{DiCarlo09} except that
$R_y^{\pi/2}$ pulse was not applied to qubit B before the tomography.
This makes the final state a superposition state. 
Also, in order to shorten the total sequence time, 
the last $R_y^{\pi/2}$ on qubit A was applied before the $U_i$ part finishes.
The real part of the final density matrices are plotted in Figs~\ref{FigS4}(b)-(e). 
No calibration for the measurement error is applied here.

%